\documentclass[aps,12pt,nofootinbib,showpacs,showkeys,preprintnumbers,amsmath,amssymb,superscriptaddress]{revtex4-1}		

\usepackage{amsmath,amsfonts,amssymb,amsthm,bm,bbm,bbold,braket,color,dsfont,fancybox,hyperref,
srcltx,subfigure,ucs,yfonts,cancel,xfrac,verbatim,xcolor}        
\usepackage{float}

\usepackage{lineno} 

\usepackage{mathtools}

\newcommand{\beq}{\begin{equation}}
\newcommand{\eeq}{\end{equation}}
\newcommand{\bea}{\begin{eqnarray}}
\newcommand{\eea}{\end{eqnarray}}
\newcommand{\beas}{\begin{eqnarray*}}
\newcommand{\eeas}{\end{eqnarray*}}
\newcommand{\bi}{\begin{itemize}}
\newcommand{\ei}{\end{itemize}}

\usepackage{makecell}

\usepackage{amsmath,amsfonts,amssymb,amsthm,bm,bbm,bbold,braket,color,dsfont,fancybox,hyperref,
srcltx,slashed,ucs,yfonts,cancel,xfrac,verbatim,xcolor}

\usepackage[inline,shortlabels]{enumitem}
\usepackage{nicefrac}
\usepackage{multirow}

\DeclareMathAlphabet{\mathpzc}{OT1}{pzc}{m}{it}
\definecolor{gold}{rgb}{1,0.8,0}
\definecolor{nara}{rgb}{1,0.4,0.1}
\definecolor{goldo}{rgb}{1,0.7,0}
\definecolor{greeno}{rgb}{0,0.8,0}
\def\bes{\begin{subequations}}
\def\ees{\end{subequations}}
\def\be{\begin{equation}}
\def\ee{\end{equation}}
\def\bea{\begin{eqnarray}}
\def\eea{\end{eqnarray}}
\def\ba{\begin{eqnarray}}
\def\ea{\end{eqnarray}}
\def\bear{\begin{array}}
\def\eear{\end{array}}

\newcommand{\bpm}{\begin{pmatrix}}
\newcommand{\epm}{\end{pmatrix}}
\newcommand{\BM}{\left(\begin{array}}		
\newcommand{\BMC}{\left[\begin{array}}		

\newcommand{\EM}{\end{array}\right)}		
\newcommand{\EMC}{\end{array}\right]}		

	% ...

\newcommand{\com}[1]{}
\newcommand{\K}{\mathcal{K}}

\begin{abstract}
In this work, we study the lepton number violating $B_{c}$ meson decays via one intermediate on-shell heavy neutrino $N$. The specific studied process is $B_{c}^{+} \to \mu^{+}  \ N \to \mu^{+} \mu^{+} \tau^{-} \nu$ which could allow distinguishing the nature of the heavy neutrino nature (Dirac or Majorana) by studying the tau lepton energy spectrum in the LHCb experiment. The result suggests that this signature could be observed in the collected data during the HL-LHCb lifetime.
\end{abstract}

\begin{document}

\title{Unveiling the Heavy Neutrino Nature at LHCb}
\author{G. A. Vasquez}
\email{g.vasquez@cern.ch {currently at the Physiks deparment of University of Zurich}}
\affiliation{Department of Physics and Astronomy, University of Victoria, Victoria BC, Canada.}
\author{Jilberto Zamora-Saa}
\email{jilberto.zamorasaa@cern.ch, jilberto.zamora@unab.cl}
\affiliation{Center for Theoretical and Experimental Particle Physics, Facultad de Ciencias Exactas, Universidad Andres Bello, Fernandez Concha 700, Santiago, Chile}
\affiliation{Millennium Institute for Subatomic physics at high energy frontier - SAPHIR, Fernandez Concha 700, Santiago, Chile.}



\maketitle

\section{Introduction}
\label{s1}
The standard model (SM) of physics is a highly successful theoretical framework that encompasses the fundamental particles and forces of nature, encompassing quarks, leptons, and bosons. However, there are various phenomena in the universe that the standard model fails to explain. These include the baryonic asymmetry of the universe (BAU), dark matter (DM), and neutrino oscillations (NOs). Over the past few decades, experiments on NOs have demonstrated that active neutrinos ($\nu$) are exceedingly light with a significant mass of around $M_{\nu} \sim 1$eV~\cite{Fukuda:1998mi,Eguchi:2002dm, PhysRevLett.87.071301}. Consequently, it is evident that the standard model is not a final theory and necessitates expansion. Among the extensions to the standard model, which provide an explanation for the minuscule masses of active neutrinos, are those rooted in the See-Saw Mechanism (SSM) \cite{Mohapatra:2005wg,Mohapatra:2006gs}. This mechanism introduces a heavy Majorana neutral lepton, commonly referred to as the Heavy Neutrino (HN), which is a singlet under the $SU(2)_L$ symmetry group. The presence of the HN ultimately leads to the existence of a very light active Majorana neutrino. These hypothetical HN's have strongly suppressed interaction with the SM particles ($Z,W^{\pm}$ bosons and $e, \mu, \tau$ leptons), doing a very tough task their detection. However, despite this suppression, the existence of HN's can be explored via rare meson decays~\cite{Dib:2000wm,Cvetic:2012hd,Cvetic:2013eza,Cvetic:2014nla,Cvetic:2015naa,Cvetic:2015ura,Dib:2014pga,Zamora-Saa:2016qlk,Milanes:2018aku,Mejia-Guisao:2017gqp,Cvetic:2020lyh}, colliders~\cite{Das:2018usr,Das:2017nvm,Das:2012ze,Antusch:2017ebe,Das:2017rsu, Das:2017zjc,Chakraborty:2018khw,Cvetic:2019shl,Antusch:2016ejd,Cottin:2018nms,Duarte:2018kiv,Drewes:2019fou,Dev:2019rxh, Cvetic:2018elt,Cvetic:2019rms,Das:2018hph,Das:2016hof,Das:2017hmg}, and tau factories \cite{Zamora-Saa:2016ito,Kim:2017pra,Dib:2019tuj}. 

A well-motivated extension of the Standard Model (SM) known as the Neutrino-Minimal-Standard-Model ($\nu$MSM)~\cite{Asaka:2005an,Asaka:2005pn} has been proposed. The $\nu$MSM is based on the Seesaw Mechanism (SSM) and introduces three heavy neutrinos. Among these, two HN's have nearly identical masses of around 1 GeV (denoted as $M_{N1}$ and $M_{N2}$), while the third one has a mass of approximately keV and is considered a candidate for dark matter. In addition to explaining the smallness of neutrino masses and neutrino oscillations, the $\nu$MSM has cosmological implications for the early universe. It can generate a slight imbalance between matter and antimatter through a phenomenon called Heavy Neutrino Oscillations (HNOs), which is also known as the Akhmedov-Rubakov-Smirnov (ARS) mechanism~\cite{Akhmedov:1998qx}.

In a previous publication \cite{Tapia:2021gne}, we provided an explanation of the impacts arising from Heavy Neutrino Oscillations (HNOs) in the rare decays of pseudoscalar $B$ mesons, specifically those violating Lepton Number (LNV) and Lepton Flavor (LFV). These decays involve two nearly indistinguishable heavy Majorana neutrinos ($M_{N_i} \sim 1$GeV), which can undergo oscillations among themselves. The objective of this article is to introduce a technique that facilitates the identification of the heavy neutrino at HL-LHCb by utilizing the remarkable detector resolution  \cite{Aaij:2021jky,Aaij:2020buf}, thereby enabling the potential observation of HNOs.

The work is arranged as follows: In Sec.~\ref{sec:production}, we described the production of heavy neutrinos mechanism in $B_c$ meson decays. In Sec.~\ref{sec:results}, we discuss the simulations of the HN production at LHCb. In Sec.~\ref{sec:summ}, we present the summary and shows the conclusions.


\section{Production of Heavy Neutrinos}
\label{sec:production}
\begin{figure}[h]
\centering
\includegraphics[scale = 0.65]{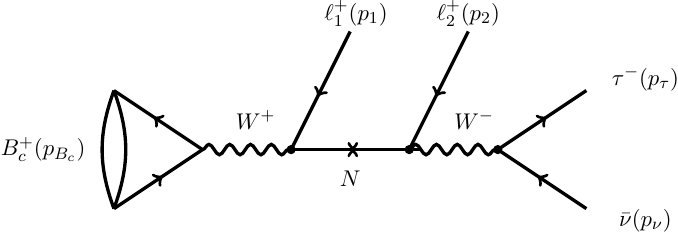}\hspace{0.3 cm}
\includegraphics[scale = 0.65]{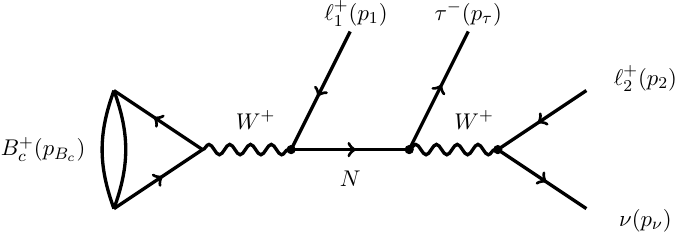}
\caption{The rare $B_{c}^{+}$ meson decay, intermediated by a heavy neutrino. Left Panel: Feynman diagrams for the LNV process $B_c^+ \rightarrow \ell^{+}_1 \ell^{+}_2 \tau^{-} \bar{\nu}$. Right Panel: Feynman diagrams for the LNC process $B_c^+ \rightarrow \ell^{+}_1 \tau^{-} \ell^{+}_2 \nu$. In this study, we will focus on a scenario where $\ell_1=\ell_2=\mu$.}
\label{fig:feynn}
\end{figure}

As we stated in the introduction, we are interested in studying the lepton number violating (LNV) and the lepton number conserving (LNC) in rare $B_c$ meson decay processes.  The LNV process can be intermediated only by Majorana HN, while the LNC by Majorana and Dirac HN (see Fig.~\ref{fig:feynn}).
The decay width for the studied processes, in terms of 4-body invariant phase space $d_{4}(B^+_c \to \mu^+ \mu^+ \tau^- \bar{\nu})$ in terms of the squared amplitude $|{\rm A^+_{\rm X}}|^2$, and for the HN kinematically allowed mass range $(M_{\tau}+M_{\mu}) \leq M_N \leq (M_{B_c}-M_{\mu})$ is
\begin{equation} 
\Gamma_{\rm X}(B^+_c \to \mu^+ \mu^+ \tau^- \bar{\nu})=\frac{1}{2M_{B_c} (2\pi)^{8}} \int d_{4}(B^+_c \to \mu^+ \mu^+ \tau^- \bar{\nu}) |{\rm A^+_{\rm X}}|^2,
\label{DBcwidth}
\end{equation}
where ${\rm X=LNC}$ and ${\rm X=LNV}$ stands for the lepton number conserving and lepton number violating processes, respectively. The squared amplitudes in terms of particles 4-momenta and the propagators are given by

\begin{flalign} 
&|{\rm A^+_{LNV}}|^2 = 256\ G_F^4\ |V_{cb}|^2\ f_{B_{cb}}^2 \ |T_{LNV}|^2\ (p_2 \cdot p_{\nu}) \Big[2 (p_1 \cdot p_{B_c}) (p_{\tau} \cdot p_{B_c})-M_{B_c}^2(p_1 \cdot p_{\tau}) \Big]&\\
\nonumber
&|{\rm A^+_{LNC}}|^2 = 256\ G_F^4\ |V_{cb}|^2\ f_{B_{cb}}^2 \ |T_{LNC}|^2\  (p_{\tau} \cdot p_{\nu}) \Biggl( \Bigg. 2 M_1^2 (p_2 \cdot p_{B_c}) \Big[M_{B_c}^2-(p_1 \cdot p_{B_c})\Big]&\\
&\quad \quad \quad \quad \times(p_1 \cdot p_2) \Big[ M_{B_c}^4- M_{B_c}^2 M_1^2+4(p_1 \cdot p_{B_c})^2-4M_{B_c}^2(p_1 \cdot p_{B_c}) \Big]  \Bigg. \Biggr),&
\label{AMPS}
\end{flalign}

where the propagators are

\begin{align} 
 T_{LNC} = \frac{B_{\tau N}B_{\mu N}^*}{P_N^2-M_N^2+i \Gamma_N^{\eta} M_N} \quad ; \quad  T_{LNV} = \frac{M_N B_{\mu N}^*B_{\mu N}^*}{P_N^2-M_N^2+i \Gamma_N^{\eta} M_N}.
\label{propa}
\end{align}

\begin{figure}[h]
\centering
\includegraphics[scale = 0.48]{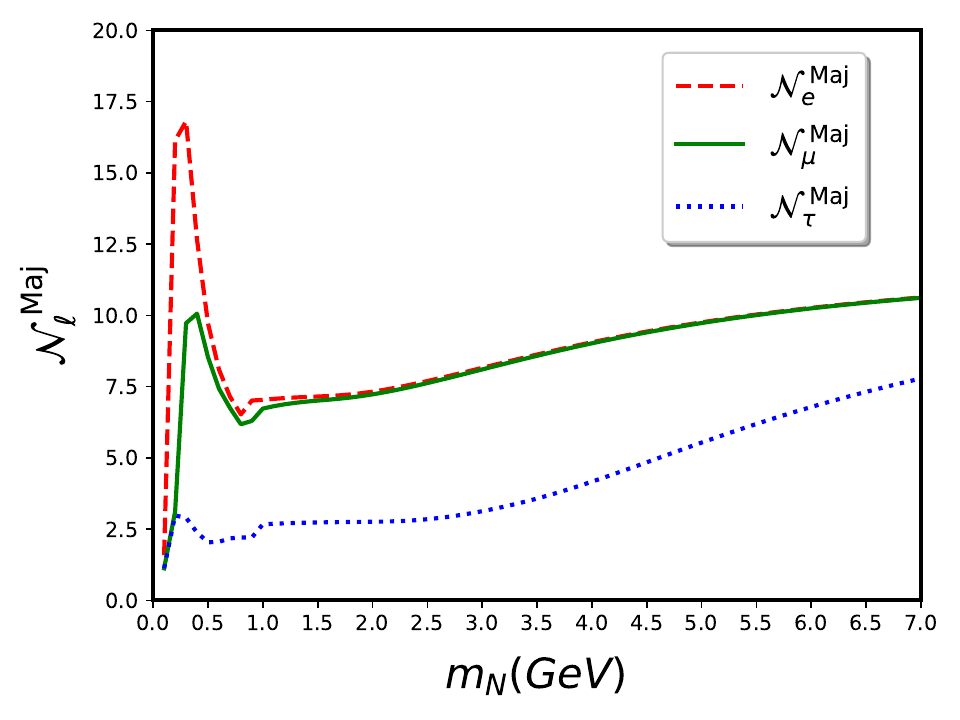}\hspace{0.1 cm}
\includegraphics[scale = 0.48]{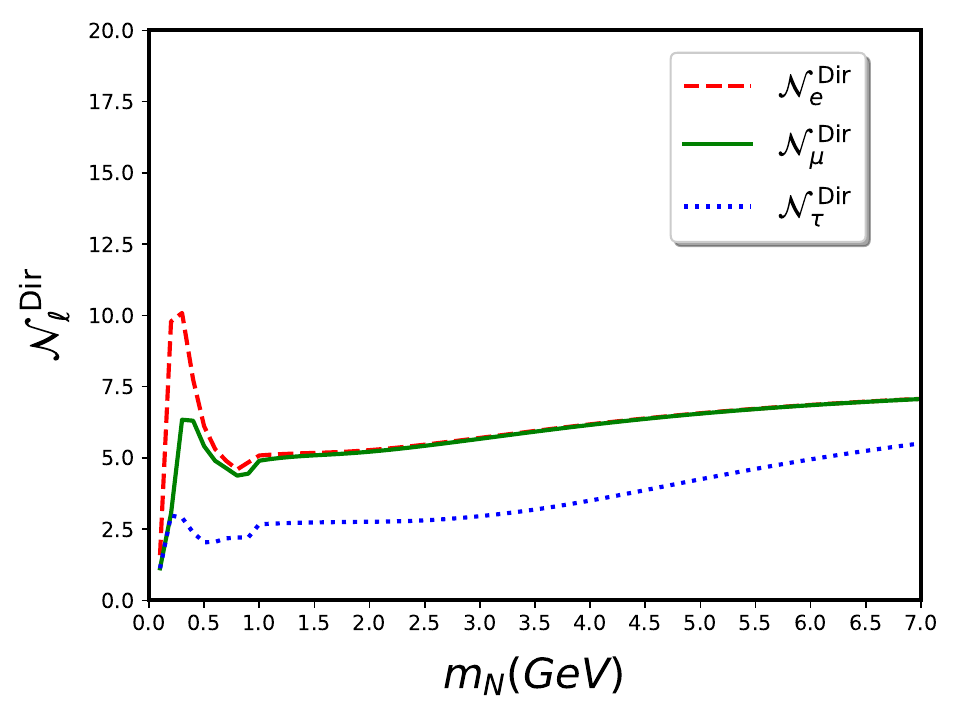}
\caption{The effective mixing coefficients ${\cal N}_{\ell}^{\eta}$. Left Panel: ${\cal N}_{\ell}^{\rm Maj}$ for Majorana heavy neutrinos. Right Panel: ${\cal N}_{\ell}^{\rm Dir}$ for Dirac heavy neutrinos.}
\label{fig:efcoef}
\end{figure}

The factors $ f_{B_{cb}}=0.322$ GeV \cite{Cvetic:2004qg} and $V_{cb}=0.041$ \cite{ParticleDataGroup:2012pjm} correspond to the decay constant and the CKM matrix element for $B_c$ meson, respectively.
In Eq.\ref{propa}, the factor  $\Gamma_N^{\eta}$ is the total heavy neutrino decay width, which in principle, can be different for Dirac ($\eta={\rm Dir}$) and Majorana ($\eta={\rm Maj}$) heavy neutrinos

\begin{equation}
\Gamma^{\eta}_{N} \equiv \Gamma^{\eta}(m_{N}) \approx  \K^{\eta}\ \frac{G_F^2 m_{N}^5}{96\pi^3}\ , 
\label{DNwidth}
\end{equation}
 here $G_F \approx 1.166 \times 10^{-5}\ {\rm GeV}^{-2}$ \cite{Mohr:2012tt} is the Fermi couplings constant. The factor $\K^{\eta}$ is given by
 \begin{equation} 
 \K^{\eta} = {\cal N}_{e}^{\eta} \; |B_{e N}|^2 + {\cal N}_{\mu}^{\eta} \; |B_{\mu N}|^2 + {\cal N}_{\tau}^{\eta} \; |B_{\tau N}|^2,
\label{DNwidth1}
\end{equation}

where the factors $B_{\ell N}$ are the heavy-light mixing elements of the PMNS matrix\footnote{In this work we define the light neutrino flavor state as $\nu_{\ell}=\sum_{i=1}^{3}U_{\ell i} \nu_i + B_{\ell N} N$. Nevertheless, other literature uses $U_{\ell N}$ or $V_{\ell N}$ as the heavy-light mixings elements  (i.e. $B_{\ell N} \equiv  U_{\ell N} \equiv  V_{\ell N}$).} which in this work are set to $|B_{e N}|^2=1 \times 10^{-8}$, $|B_{\mu N}|^2=5 \times 10^{-7}$ and $|B_{\tau N}|^2=5 \times 10^{-6}$ all of these widely allowed by current limits~\cite{Boiarska:2021yho,Abdullahi:2022jlv}, the factors ${\cal N}_{\ell}^{\eta}$ are the effective mixing coefficients which account for all possible decay channels of $N$ and are presented in Fig.~\ref{fig:efcoef} for our HN mass of interest ($0 \leq m_N \leq 7.0$ GeV).

\begin{figure}[htb]
\centering
\includegraphics[scale=0.65]{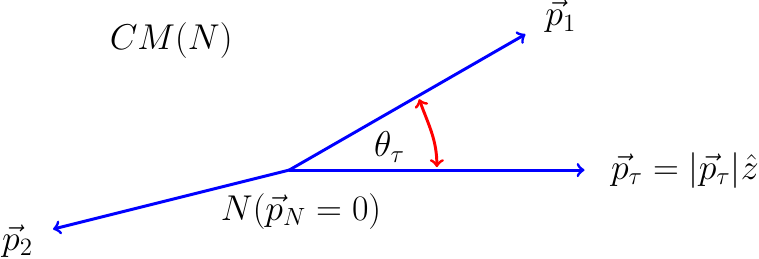}
\caption{Scheme of the final 3-momenta. Here for simplicity, we have omitted the vectors $\vec{p}_{\nu}$ and $\vec{p}_{B_c}$, however, we have represented the relevant angle $\theta_{\tau}$.}
\label{fig:angles}
\end{figure}

Due to the different 4-momenta structure between $|{\rm A^+_{LNV}}|^2$ and $|{\rm A^+_{LNV}}|^2$ (see Eq. \eqref{AMPS}), it is possible to infer that the energy spectra of the final tau lepton is an appropriate variable to distinguish the HN nature (see references \cite{Cvetic:2012hd} and \cite{Cvetic:2015naa} for a detailed discussion). The $\tau$ lepton energy spectra for the LNV process is given by
\bea
&& \frac{d Br^{(LNV)} (B_c^+ \to \ell_1^+ \ell_2^+ \tau^- \bar{\nu})}{d E_{\tau} d \cos \theta_{\tau}} =
\frac{Z^{(LNV)}}{\Gamma(B_c \to {\rm all})} \frac{\left[ m_N (m_N - 2 E_{\tau}) + m_{\tau}^2 - m_2^2 \right]^2}{
4 m_N \left[  m_N (m_N- 2 E_{\tau}) + m_{\tau}^2 \right]}
\nonumber\\
&& \times
{\bigg \{}
\cos \theta_{\tau}
(m_N^2 -m_1^2) \left[
 (m_{B_c}^2 - m_N^2)^2 - 2 m_1^2 (m_{B_c}^2 + m_N^2) + m_1^4 \right]^{1/2}
 (E_{\tau}^2 - m_{\tau}^2)
\nonumber\\
&& +
\left[ m_N^2 (m_{B_c}^2 - m_N^2) + m_1^2 (m_{B_c}^2 + 2 m_N^2) - m_1^4
\right] E_{\tau} \sqrt{E_{\tau}^2-m_{\tau}^2}
{\bigg \}}
+ (m_1 \leftrightarrow m_2) ,
\label{dGM}
\eea
where the angle $ \theta_{\tau}$ is define as the angle between $\vec{p_1}$ and $\vec{p_\tau}$ (see Fig.~\ref{fig:angles}), the function $Z^{(LNV)}$ is defined as
\be
Z^{(LNV)} \equiv  \left( 1 - \frac{1}{2} \delta_{\ell_1,\ell_2} \right)
 G_F^4 f_{B_c}^2 |B_{\ell_1 N}^{*} B_{\ell_2 N}^{*} V_{cb}^{*}|^2
\frac{2}{(2 \pi)^4} \frac{m_N }{ \Gamma_N^{\rm Maj} m_{B_c}^3}
 \lambda^{1/2}(m_{B_c}^2, m_N^2, m_1^2) \ ,
\label{ZM}
\ee
and $\lambda^{1/2}$ is the square root of the function
\be
\lambda(x,y,z) \equiv x^2 + y^2 + z^2 - 2 xy - 2 yz - 2 zx .
\label{lamdef}
\ee

It is important to remark that the factor $\left( 1 - \frac{1}{2} \delta_{\ell_1,\ell_2} \right)$ accounts for the case when $\ell_1 \neq \ell_2$, however in our case $\ell_1 = \ell_2 = \mu$, then $\left( 1 - \frac{1}{2} \delta_{\mu,\mu} = 1/2 \right)$. On the other hand, the factor $(m_1 \leftrightarrow m_2)$ accounts for the case, when $\ell_1$ is produced at $\ell_2$ vertex and vice-versa (crossed channel).
The integration over the angle $\theta_{\tau}$ gives
\bea
&&\frac{d Br^{(LNV)}}{d E_{\tau}}
({B_c}^+ \to \ell_1^+ \ell_2^+ \tau^- {\bar \nu} )
=  \frac{Z^{(LNV)}}{\Gamma(B_c \to {\rm all})} \frac{1}{2 m_N}
\left[ m_{B_c}^2 (m_N^2 + m_1^2) - (m_N^2 - m_1^2)^2 \right]
\nonumber\\
&& \times
E_{\ell} \sqrt{E_{\tau}^2 - m_{\tau}^2}
\frac{\left(m_N^2 - 2 m_N E_{\tau} + m_{\tau}^2 - m_2^2 \right)^2}
{ \left(m_N^2 - 2 m_N E_{\tau} + m_{\tau}^2 \right)}
+ (m_1 \leftrightarrow m_2),
\label{dGMEel}
\eea
for the lepton number conserving (LNC) processes we have
 \bea
&& \frac{d Br^{(LNC)}}{d E_{\tau} d \cos \theta_{\tau}} =
\frac{Z^{(LNC)}}{\Gamma(B_c \to {\rm all})}
\frac{(-1) \sqrt{E_{\tau}^2-m_{\tau}^2}  \left[ -m_2^2 + m_{\tau}^2 +
     m_N (m_N - 2 E_{\tau}) \right]^2}{24 m_N \left[m_{\tau}^2 +
     m_N (m_N - 2 E_{\tau}) \right]^3}
\nonumber\\ &&
\times {\bigg \{} \cos \theta_{\tau} (m_1^2 - m_N^2)  \sqrt{E_{\tau}^2-m_{\tau}^2}
\sqrt{((m_{B_c}+m_1)^2 - m_N^2) ((m_{B_c}-m_1)^2 - m_N^2)}
\nonumber\\ &&
\times
{\big [}
\left( 3 m_{\tau}^2 + m_N (m_N - 4 E_{\tau}) \right)
\left(   m_{\tau}^2 + m_N (m_N - 2 E_{\tau}) \right)
+ m_2^2
\left( 3 m_{\tau}^2 - m_N (m_N + 2 E_{\tau}) \right)
{\big ]}
\nonumber\\ &&
 +
{\big [}
\left( m_1^4 - m_N^2 (m_{B_c}^2 - m_N^2) - m_1^2 (m_{B_c}^2 + 2 m_N^2) \right)
{\big (}
8 E_{\tau}^3 m_N^2 - 2 m_{\tau}^2 m_N (2 m_2^2 + m_{\tau}^2 + m_N^2)
\nonumber\\ &&
+ 2 E_{\tau}^2 m_N (m_2^2 + 5 m_{\tau}^2 + 5 m_N^2)
 + E_{\tau} ( 3 m_2^2 m_{\tau}^2 + 3 m_2^2 m_N^2 +
(3 m_{\tau}^2 + m_N^2) (m_{\tau}^2 + 3 m_N^2)) {\big )}
{\big ]}
{\bigg \}}
\nonumber\\
&& + (m_1 \leftrightarrow m_2),
\label{dGdEdtheta_LNC}
\eea
where $Z^{(LNC)}$ is defined as
\be
Z^{(LNC)} \equiv  G_F^4 f_{B_c}^2
|B_{\ell_1 N}^{*} B_{\tau N}  V_{cb}^{*}|^2
\left( 1 - \frac{1}{2} \delta_{\ell_1,\ell_2} \right)
\frac{2}{(2 \pi)^4} \frac{ m_N }{\Gamma_N^{\rm Dir} m_{B_c}^3}
 \lambda^{1/2}(m_{B_c}^2, m_N^2, m_1^2) .
\label{ZD}
\ee
The integration over $\theta_{\tau}$ gives
\bea
\lefteqn{
\frac{d Br^{(LNC)}}{d E_{\tau}}({B_c}^+ \to \ell_1^+ \ell_2^+ \tau^- {\nu} )
=  \frac{Z^{(LNC)}}{\Gamma(B_c \to {\rm all})} \frac{1}{96 m_N^2}
\frac{1}{\left[ m_{\tau}^2 + m_N (-2 E_{\tau} + m_N) \right]^3}}
\nonumber\\
&&  \times
{\bigg \{} 8 \sqrt{(E_{\tau}^2 - m_{\tau}^2)}
   m_N \left[ m_2^2 - m_{\tau}^2 + (2 E_{\tau} - m_N) m_N \right]^2
\nonumber\\
&& \times
\left[ -m_1^4 + m_{B_c}^2 m_N^2 - m_N^4 + m_1^2 (m_{B_c}^2 + 2 m_N^2) \right]
{\big [} 8 E_{\tau}^3 m_N^2 - 2 m_{\tau}^2 m_N (2 m_2^2 + m_{\tau}^2 + m_N^2)
\nonumber\\
&&
- 2 E_{\tau}^2 m_N \left( m_2^2 + 5 (m_{\tau}^2 + m_N^2) \right)
+ E_{\tau} (3 m_{\tau}^4 + 10 m_{\tau}^2 m_N^2 + 3 m_N^4 +
3 m_2^2 (m_{\tau}^2 + m_N^2))
{\big ]}
{\bigg \}}
\nonumber\\
&&
+ (m_1 \leftrightarrow m_2 ) \ ,
\label{dGdE_LNC}
\eea

An important suppression effect acting on the decay width comes from the finite detector length ($L_D$), this effect is named Acceptance Factor ($\rm AF^{\eta}$) and can be written as
\begin{equation}
    {\rm AF^{\eta}}=1-e^{\frac{L_D \Gamma_N^{\eta}}{\gamma_{N} \beta_{N}}},
\end{equation}
where ${\rm \eta = Dir, Maj}$, the factor $\gamma_N$ stands for the HN Lorentz factor, and $\beta_N$ for the HN velocity, in our analysis, we will use $\gamma_N \beta_N = 2$ (see appendix \ref{app1} for more details) and $|B_{e N}|^2=5 \times 10^{-7}$  and $|B_{\tau N}|^2=5 \times 10^{-6}$, which are not excluded for current limits~\cite{Boiarska:2021yho,Abdullahi:2022jlv}. Therefore, the effective (real) branching ratio can be written as follow
\begin{subequations}  
\begin{align}
{\rm Br_{eff}^{Maj}} &= \epsilon \times \Big( 1-e^{\frac{L_D \Gamma_N^{\rm Maj}}{\gamma_{N} \beta_{N}}}\Big)\ \times \ \frac{{\rm Br_{eff}^{LNC}}+{\rm Br_{eff}^{LNV}}}{\rm \Gamma_N^{Maj}} \equiv \epsilon \times {\rm AF^{Maj}}\ \times \ \Bigg( \frac{{\rm Br_{eff}^{LNC}}+{\rm Br_{eff}^{LNV}}}{\rm \Gamma_N^{Maj}}\Bigg),
\\
{\rm Br_{eff}^{Dir}} &= \epsilon \times \Big(1-e^{\frac{L_D \Gamma_N^{{\rm Dir}}}{\gamma_{N} \beta_{N}}}\Big)\ \times \ \frac{{\rm Br_{eff}^{LNC}}}{\rm \Gamma_N^{Dir}} \equiv \epsilon \times {\rm AF^{Dir}}\ \times \ \frac{{\rm Br_{eff}^{LNC}}}{\rm \Gamma_N^{Dir}}.
\end{align}
\label{Breffefctive}
\end{subequations}  
Here we have introduced the factor $\epsilon$ which will account for the detector efficiency. We remark while for Majorana HN both channels (LNC and LNV) contribute, for Dirac ones only the LNC channel does.


\section{Results}
\label{sec:results}

In this section, we will present the results obtained through the simulations and Eqs.~\eqref{Breffefctive}. We emphasize that  $\gamma_N \beta_N = 2$ has been used in all the results.

\begin{figure}[H]
\centering
\includegraphics[scale = 0.5]{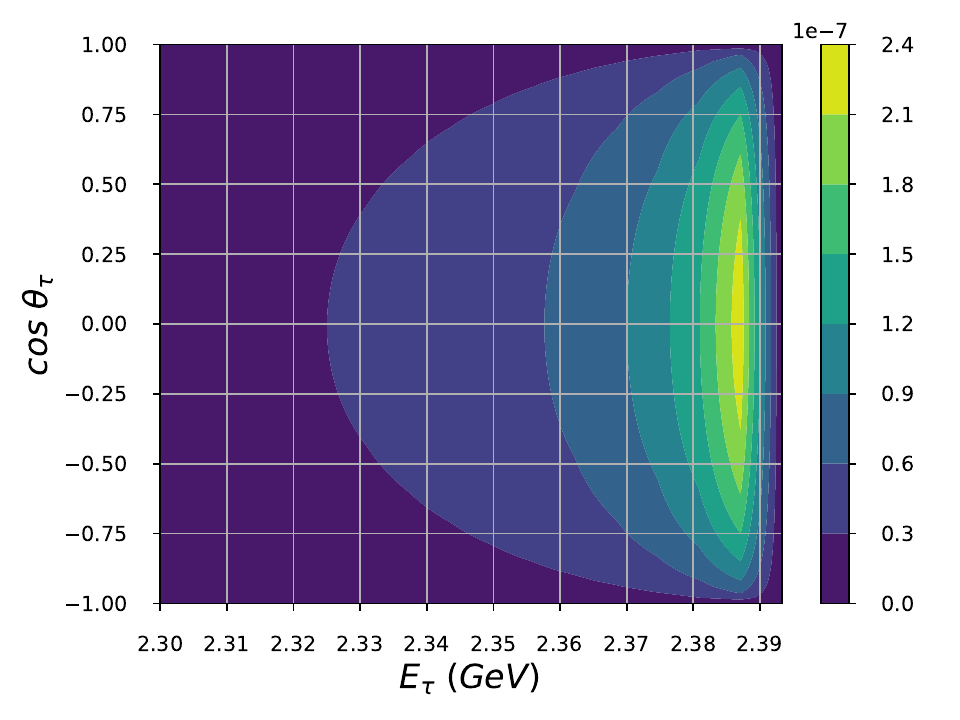}\hspace{0.01cm}
\includegraphics[scale = 0.5]{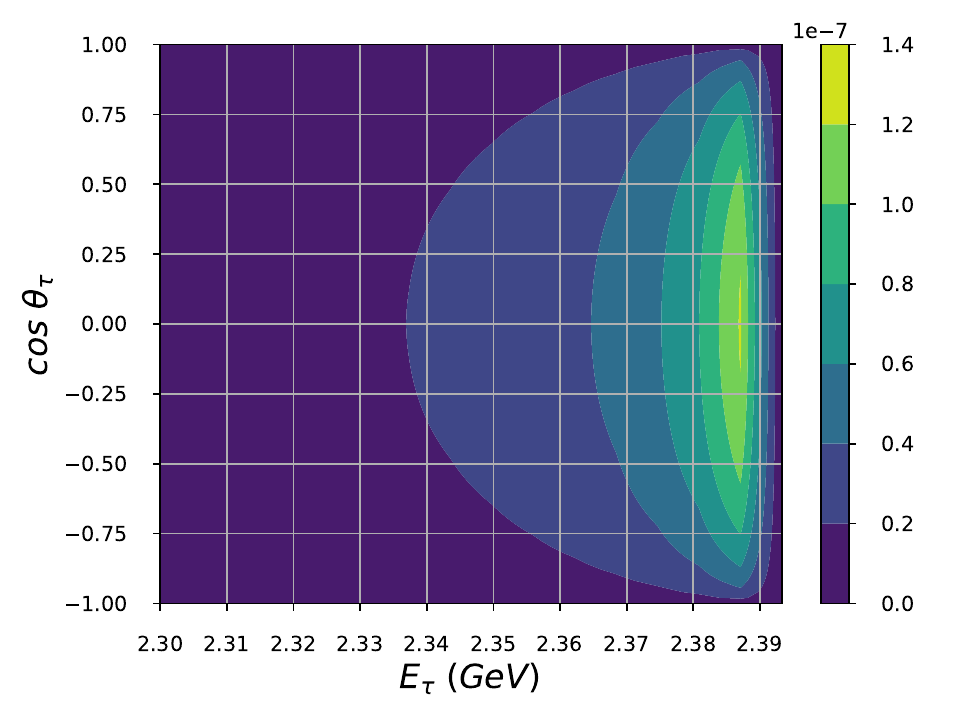}
\caption{Branching ratio distribution $d Br^{(X)} / d E_{\tau} d \cos \theta_{\tau}$. Left panel $X=$ Dirac and Right panel $X=$ Majorana. Here $M_N = 4.0$ GeV, $\epsilon = 1.0$, $|B_{\mu N}|^2=5 \times 10^{-7}$ and $|B_{\tau N}|^2=5 \times 10^{-6}$}
\label{fig:dBrdEdtheta}
\end{figure}

The Fig.~\ref{fig:dBrdEdtheta} left panel (Dirac HN) shows that the branching ratio distribution is maximum $\frac{d Br^{\rm (Dir)}}{d E_{\tau} d\cos \theta_{\tau}}\approx 2.27 \times 10^{-7}$ $GeV^{-1}$ for  $E_{\tau} \approx 2.385$ GeV and $-0.38 \leq \cos(\theta_{\tau}) \leq 0.38$. In the case of Fig.~\ref{fig:dBrdEdtheta} right panel (Majorana HN) the branching ratio distribution is maximum $\frac{d Br^{\rm (Maj)}}{d E_{\tau} d\cos \theta_{\tau}}\approx 1.22 \times 10^{-7}$ $GeV^{-1}$ for $ E_{\tau} \approx 2.387 $ GeV and $-0.17 \leq \cos(\theta_{\tau}) \leq 0.17$.

\begin{figure}[h]
\centering
\includegraphics[scale = 0.5]{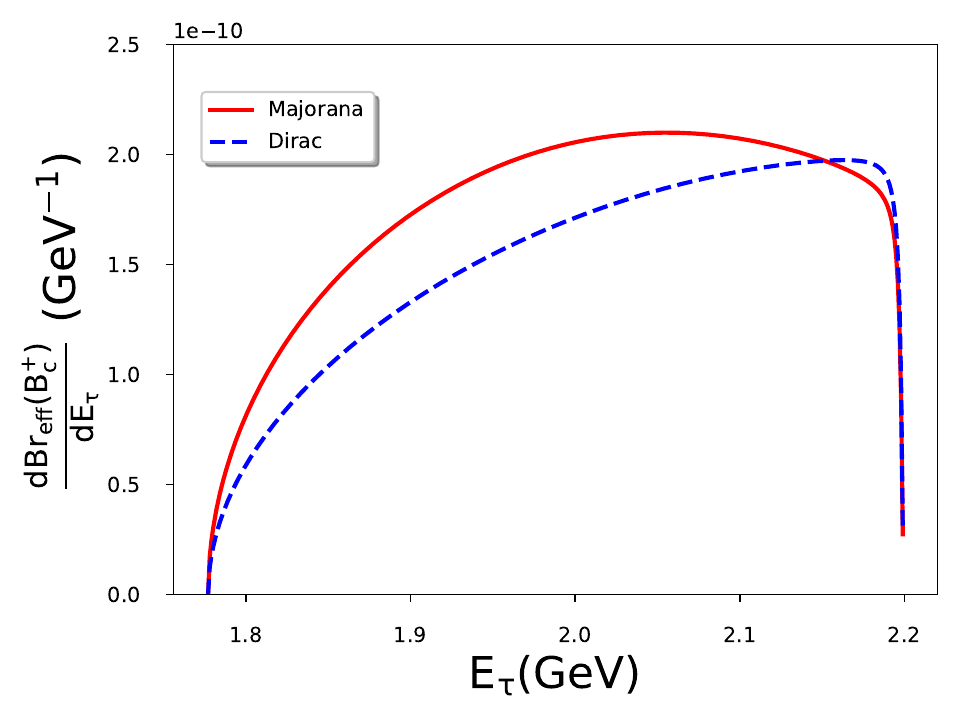}\hspace{0.01cm}
\includegraphics[scale = 0.5]{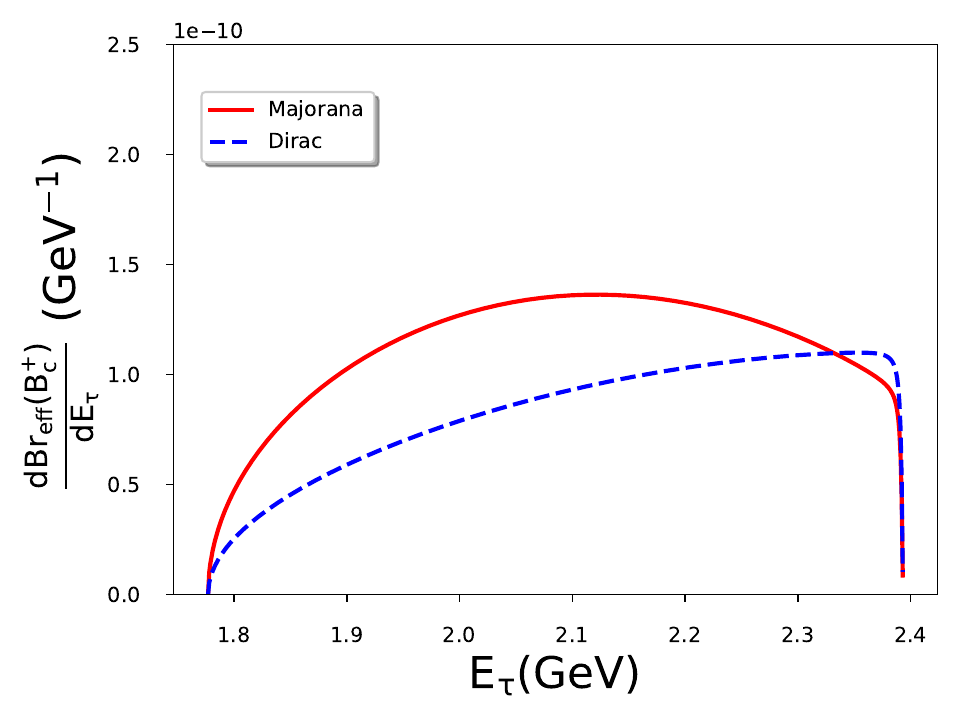}
\caption{Branching ratio distribution $d Br / d E_{\tau}$. Left panel: $M_N = 3.5$ GeV and Right panel $M_N = 4.0$ GeV. Here $\epsilon = 1.0$, $|B_{\mu N}|^2=5 \times 10^{-7}$ and $|B_{\tau N}|^2=5 \times 10^{-6}$.}
\label{fig:dBrdE}
\end{figure}

In Fig.~\ref{fig:dBrdE} two mass cases are presented to illustrate the behavior of the branching ratio distribution $\rm d Br_{eff}/d E_{\tau}$. In the left panel ($M_N = 3.5$ GeV) is possible to observe that in the range $ 1.77 \leq M_N \leq 2.15$ GeV of the energy the Majorana case dominates over Dirac. On the contrary in the $ 2.15 \leq M_N \leq 2.19$ GeV range the Dirac dominates over Majorana. Similarly, for Neutrino masses of 4.0 GeV (right panel), the Majarona dominates in the $ 1.77 \leq M_N \leq 2.33$ GeV HN mass range and Dirac on the $ 2.33 \leq M_N \leq 2.39$ GeV range, however, the difference in the slope between Majorana and Dirac is more evident. 

\begin{figure}[hbt]
\centering
\includegraphics[scale = 1.0]{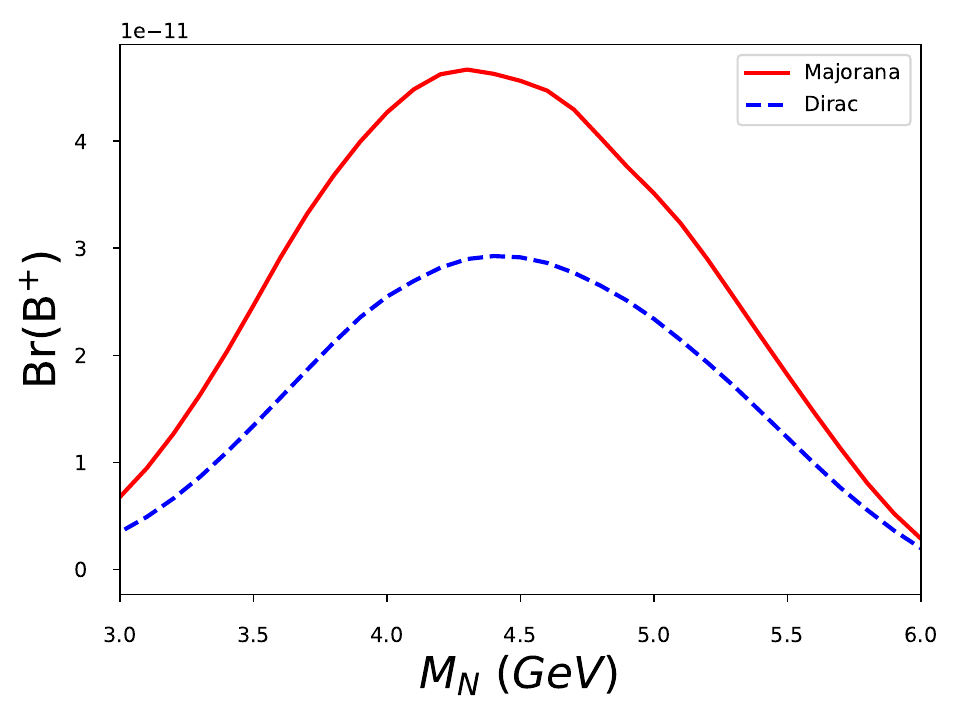}
\caption{Heavy neutrino effective branching ratios. The red solid line stands for the Majorana case, and the dashed blue line for the Dirac case. Here we have used $\epsilon=0.8$,  $|B_{e N}|^2=5 \times 10^{-7}$  and $|B_{\tau N}|^2=5 \times 10^{-6}$.}
\label{fig:effBranRat}
\end{figure}

In order to estimate a realistic number of HN that can be produced at the HL-LHCb, we will consider the detector efficiency $\epsilon=0.8$, which is under a conservative approach~\cite{LHCb:2014nio}. Therefore, 
in Fig.~\ref{fig:effBranRat} we present the values of effective branching ratios (Eqs.~\ref{Breffefctive}) over our range of interest for heavy neutrino masses ($3 \lesssim M_N \lesssim 6$ GeV) for the above mention efficiency. 
On the other hand, considering a luminosity of about ${\cal L} = 10^{34}\ {\rm cm^{-2} \ sec^{-1}}$, one could expect the total amount of $B_c$ mesons produced of the order of $N_{B_c} \sim 5 \times 10^{10}$ per year~\cite{Gouz_2004}. In Table~\ref{NN} we show the expected number of HN $N_N^{X}$ (X=Dir/Maj) for the two HN studied masses ($M_N = 3.5, 4.0$), and for the two HN nature (Dirac and Majorana).

\begin{center}
\begin{table}[h]
\centering
\begin{tabular}{|c|c|c|c|c|c|}
\hline
\ \ $ M_N$ GeV \ \ &\ \ \thead{Operation time \\ (years)} \ \ & $Br^{\rm Dir}$ & \ \ $N_N^{\rm Dir}$ \ \ &\ \  $Br^{\rm Maj}$\ \ &\ \ $N_N^{\rm Maj}$\ \  \\ \hline
3.5  &     5                   &     $1.34 \times 10^{-11}$     &    $\approx 3$      &  $2.47 \times 10^{-11}$           &      $\approx 6$       \\ \hline
3.5  &     10                  &     $1.34 \times 10^{-11}$     &    $\approx 7$      &   $2.47 \times 10^{-11}$          &     $\approx 12$        \\ \hline
3.5  &     15                  &     $1.34 \times 10^{-11}$     &    $\approx 11$      &   $2.47 \times 10^{-11}$          &     $\approx 19$        \\ \hline
4.0  &     5                   &     $2.55 \times 10^{-11}$    &     $\approx 6$      &   $4.27 \times 10^{-11}$          &     $\approx 11$        \\ \hline
4.0  &     10                  &     $2.55 \times 10^{-11}$     &    $\approx 13$      &  $4.27 \times 10^{-11}$          &    $\approx 21$         \\ \hline
4.0  &     15                  &     $2.55 \times 10^{-11}$     &    $\approx 19$      &  $4.27 \times 10^{-11}$          &    $\approx 32$        \\ \hline
\end{tabular}
\caption{Expected number of HN at HL-LHCb with an overall detector efficiency of 0.8. Here we have used $|B_{\mu N}|^2=5 \times 10^{-7}$ and $|B_{\tau N}|^2=5 \times 10^{-6}$.}
\label{NN}
\end{table}
\end{center}

\section{Summary and conclusions}\label{sec:summ}

In this work, we have studied the production of HN's via the rare $B_c$ meson decay $B_{c}^{+} \to \mu^{+}  \ N \to \mu^{+} \mu^{+} \tau^{-} \nu$ in the HL-LHCb experiment. We have shown that for mixings elements $|B_{\mu N}|^2=5 \times 10^{-7}$  and $|B_{\tau N}|^2=5 \times 10^{-6}$ and for HN masses $M_N=3.5$ and $M_N=4.0 GeV$ would be possible to probe the existence of HN during the LHC-LHCb lifetime. It is worth mentioning, that we focus on a scenario with conservative values for HN mixing elements $|B_{\ell N}|$, however, there are scenarios where the HN mixings elements are less tighten $|B_{\mu N}|^2 \sim 10^{-6}$ and $|B_{\tau N}|^2 \sim 10^{-5}$ which allows producing up to 3000 HN events for Majorana case~\cite{Tapia:2021gne}. Furthermore, we emphasize that due to the different energy distributions of the final tau lepton (Fig.~\ref{fig:dBrdE}), it could be possible to reveal the HN's nature. In addition, the angular distribution (Fig.~\ref{fig:dBrdEdtheta}) between final leptons might be the key to improving the signature of the events and unveiling the Dirac and Majorana cases. 

\section{Acknowledgments}
The work of J.Z-S. was funded by ANID-Millennium Science Initiative Program - ICN2019\_044. The work of G.V. is supported by the Natural Science and Engineering Research Council, Canada.
 \appendix
 \section{Appendix I}
 \label{app1}
 
\begin{figure}[hbt]
\centering
\includegraphics[scale = 0.8]{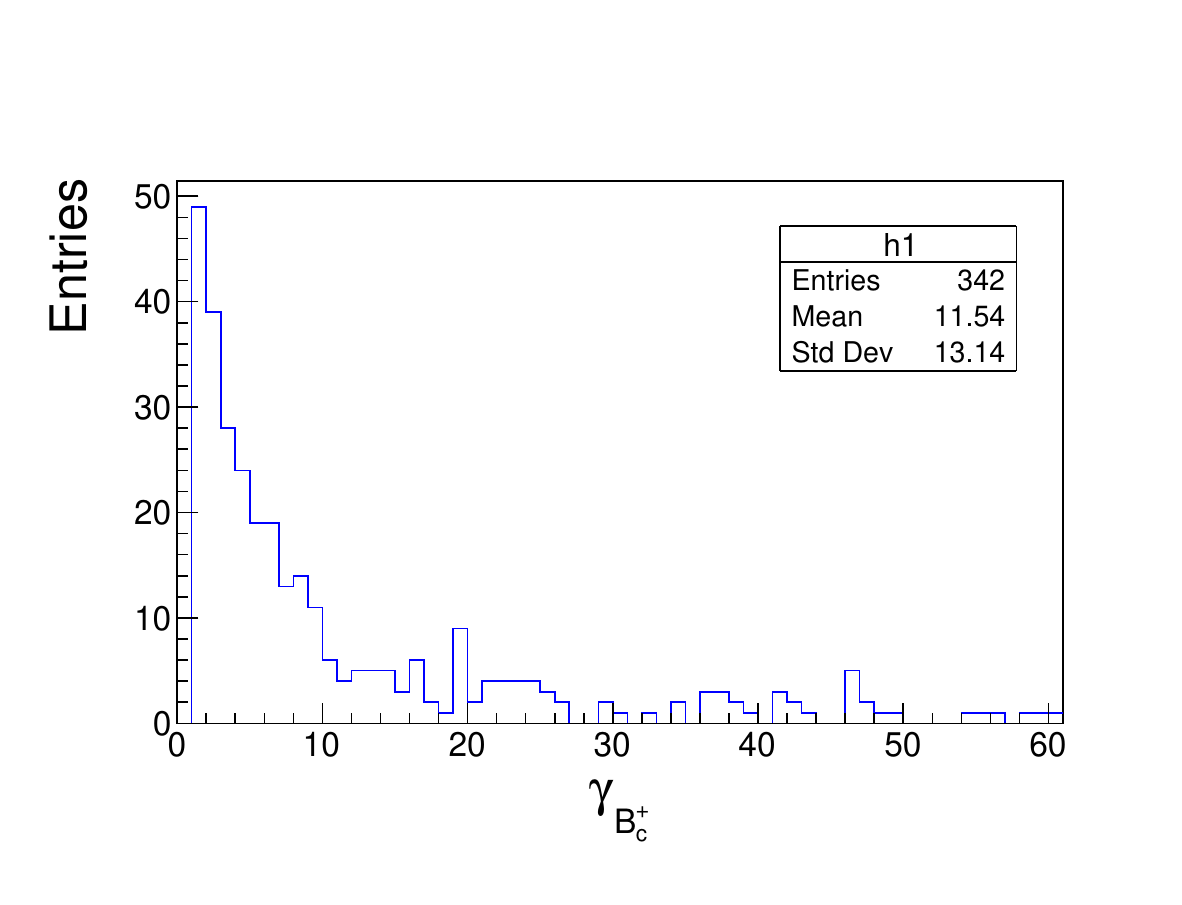}
\caption{The  $\gamma_{B^+_c}$ distribution. We notice that the most representative value is $\gamma_{B^+_c}=2$.}
\label{fig:Bcgammadistro}
\end{figure}

An appropriate evaluation of  Eqs.~\ref{Breffefctive} requires a realistic value of $\gamma_N \beta_N$, which can be obtained  from the $\gamma_{B^+_c}$ distribution by means of Lorentz transformation.  The  $\gamma_{B^+_c}$ distribution is presented in Fig.\ref{fig:Bcgammadistro}  and was obtained carrying out simulations of $B^+_c$ mesons production via charged current Drell-Yan process, using 
 \textsc{MadGraph5\_aMC@NLO}~\cite{Alwall:2014hca}, \textsc{Pythia8}~\cite{Sjostrand:2007gs} and \textsc{Delphes}~\cite{deFavereau:2013fsa}, for the LHCb conditions at $\sqrt{s}=13$ TeV. The $B^+_c$ meson velocity ($\equiv \beta_{B^+_c}$) can be obtained from $\gamma_{B^+_c}$ using $\beta_{B^+_c}=\sqrt{1-1/\gamma_{B^+_c}^2}$. The Fig. ~\ref{fig:Bcgammadistro} show the $\gamma_{B^+_c}$ distribution
 
It is worth mentioning, that, in general, $B^+_c$ is moving in the lab frame when it decays into $N$ and $\ell_1$, therefore, the product $\gamma_N \beta_N$ is not always fixed and can be written as
\begin{equation}
\beta_N \gamma_N = \sqrt{(E_N({\hat p}'_N)/M_N)^2 - 1},
\label{bNgN}
\end{equation}
where $E_N$ is the heavy neutrino energy in the lab frame, depending on the ${\hat p}'_N$ direction  in the $B^+_c$-rest frame ($\Sigma'$).

\begin{figure}[hbt]
\centering
\includegraphics[width=8cm]{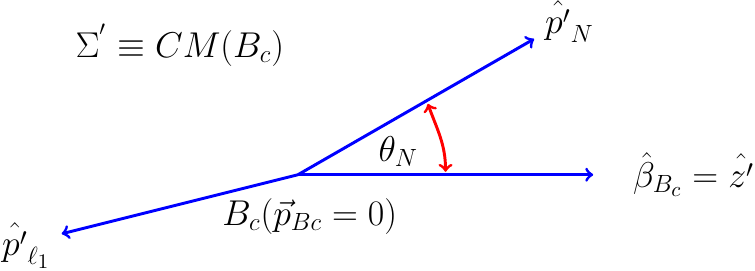}
\caption{Schematics representation of the directions of 3-momentum in the $B^+_c$-rest frame ($\Sigma'$). Here $\theta_N$ define the angle between ${\hat \beta}_{B^+_c}$ and ${\hat p}'_N$, where ${\hat \beta}_{B^+_c}=\frac{\vec{\beta}_{B^+_c}}{|\vec{\beta}_{B^+_c}|}$ is the direction of the velocity of $B^+_c$ in the lab frame, we notice that ${\hat \beta}_{B^+_c}$ also defines the ${\hat z}'$-axis.}
\label{fig:FigpN}
\end{figure}
The relation among $E_N$, $\vec{p'}_N$ and the angle $\theta_N$ is given by the Lorentz energy transformation (see Fig.~\ref{fig:FigpN})

\begin{equation}
E_N = \gamma_{B^+_c} (E'_N + \cos \theta_N \beta_{B^+_c} |{\vec p}'_N|),
\label{EN}
\end{equation}
where the corresponding factors in the $B_c$-rest frame ($\Sigma'$) are given by
\begin{equation}
E'_N = \frac{M_{Bc}^2 + M_N^2 - M_{\ell_1}^2}{2 M_{Bc}}, \quad
|\vec{p'}_N| = \frac{1}{2} M_{Bc} \lambda^{1/2} \left( 1, \frac{M_{\ell_1}^2}{M_{Bc}^2}, \frac{M_N^2}{M_{Bc}^2} \right),
\label{ENppNp}
\end{equation}
we remarks that $\beta_{Bc}$ is the velocity of ${B_c}$ in the lab frame, and $\lambda(x,y,z)$ is 
\begin{equation}
\lambda(x,y,z)=x^2+y^2+z^2-2xy-2xz-2yz
\end{equation}
 
 The $\gamma_N \beta_N$ values can range between the values presented in Table ~\ref{tab:tab1}
 
\begin{table}[hbt]
\begin{tabular}{|c|c|c|c|c|}
\hline
$M_N$ (GeV)  & $\quad \theta_N \ {\rm (rad)}\quad $  &\quad $\gamma_{N} \beta_{N}\quad $ \\ \hline
 3.5   & 0        &  3.12  \\ \hline
 3.5   & $\pi/2$  &  2.13  \\ \hline
 3.5   & $\pi$    &  1.01  \\ \hline
 4.0   & 0        &  2.73  \\ \hline
 4.0   & $\pi/2$  &  1.97  \\ \hline
 4.0   & $\pi$    &  1.13  \\ \hline
\end{tabular}
\caption{The values of $\gamma_{N} \beta_{N}$ for $\gamma_{B^+_c} = 2.0$, $\beta_{B^+_c}=0.75$ and different angles $\theta_N$.  The average value for $M_N=3.5$ GeV is $\gamma_{N} \beta_{N} = 2.01$, while for $M_N=4.0$ GeV is $\gamma_{N} \beta_{N} = 1.94$.}
\label{tab:tab1}
\end{table}

In order to perform the calculation in a simple way, during the development of this work we have considered $\gamma_{N} \beta_{N} = 2.0$, however, we stress that the result does not change significantly in the range $ 1.01 \leq \gamma_{N} \beta_{N} \leq 3.12$.

\bibliographystyle{apsrev4-1}
\bibliography{biblio.bib}
\leavevmode

\end{document}